\documentclass[12pt]{article}
\usepackage{graphicx}
\begin{document}

\title{Role of Orbitals in the Physics of Correlated Electron Systems}

\author{D.I.Khomskii\\
II.~Physikalisches Institut,\\Universit\"at zu K\"oln,\\Z\"ulpicher Str.~77,\\
50937 K\"oln, Germany}

\maketitle

Rich properties of systems with strongly correlated electrons,
such as transition metal (TM) oxides, is largely connected
with an interplay of different degrees of freedom in them: charge,
spin, orbital ones as well as crystal lattice. Specific and often
very important role is played by orbital degrees of freedom. They
can lead to a formation of different superstructures (an orbital
ordering) which are associated with particular types of structural
phase transitions --- one of very few examples where the
microscopic origin of these transitions is really known; they
largely determine the character of magnetic exchange and the type
of magnetic ordering; they can also strongly influence many other
important phenomena such as insulator-metal transitions (IMT), etc.

In this comment I will try to shortly summarize the main concepts
and discuss some of the well-known manifestations of orbital
degrees of freedom, but will mostly concentrate on a more recent
development in this field. More traditional material is covered in
several review articles \cite{Kugel-Khomskii,Tokura-Nagaosa,van
den Brink}. Although I tried to cover the main new development in
this area, the choice of topics of course is influenced by my own
interests; other people probably would have stressed other
parts of this big field.

\section{Basic notions}

\begin{figure}[htbp]
\centering
\includegraphics[width=5cm]{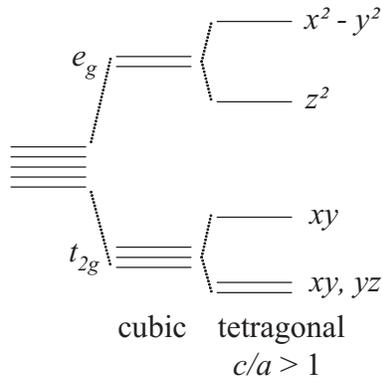}
\caption{\label{fig:levels} Schematic form of the crystal field
splitting of $d$-levels of transition metal in octahedral
coordination.}
\end{figure}

\begin{figure}[htbp]
\centering
\includegraphics[width=10cm]{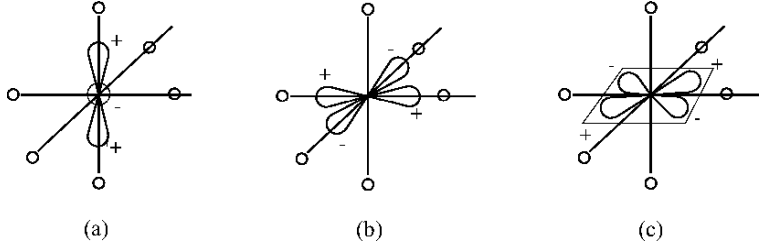}
\caption{\label{fig:orbitals} Typical shape of different orbitals:
(a) $z^2=3z^2-r^2$-orbital; (b) $x^2-y^2$-orbital; (c)
$xy$-orbital.}
\end{figure}

Five-fold degenerate $d$-levels of TM ions ($l=2$, $2l+1=5$) are
split in cubic crystal field (CF), typical for many TM compounds,
into a triply-degenerate $t_{2g}$ levels (orbitals $xy$, $xz$ and
$yz$) and doubly-degenerate $e_g$ ones ($z^2=3z^2-r^2$ and
$x^2-y^2$ orbitals), see fig.~\ref{fig:levels}. Further lowering of
CF to a tetragonal or orthorhombic one splits both $t_{2g}$ and
$e_g$ levels, whereas a trigonal (rhombohedral) distortion splits
only $t_{2g}$ levels. The shape of corresponding electron wave
functions is shown in fig.~\ref{fig:orbitals}; the notation
($x^2-y^2$, $xy$, etc.)\ actually describes the shape of
electronic density of a corresponding orbital. Of course, also
linear combinations of the basic orbitals are possible, e.g.\
\begin{equation}
|\theta\rangle = \cos(\theta/2)|z^2\rangle + \sin(\theta/2)|x^2 -
y^2\rangle\,. \label{eq1}
\end{equation}

\begin{figure}[htbp]
\centering
\includegraphics[width=5cm]{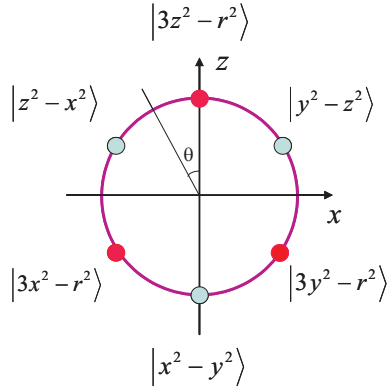}
\caption{\label{fig:circle} Different orbitals, described by
Eq.~(\ref{eq1}), in $T^x-T^z$-plane.}
\end{figure}

Such states, relevant for doubly-degenerate $e_g$ orbitals, can be
conveniently described by the pseudospin $T=\frac12$  ($T^z=\frac12$
for the $z^2$ orbital and $T^z=-\frac12$ --- for the $x^2-y^2$
one) and represented on a diagram of fig.~3. Note that in cubic CF
the axes $x$, $y$ and $z$ are equivalent, i.e.\ the orbital
$|z^2\rangle = |3z^2 - r^2\rangle$ should be equivalent to
$|x^2\rangle = |3x^2 - r^2\rangle$ and $|y^2\rangle = |3y^2 -
r^2\rangle$. These later ones correspond to the angles $\theta =
\pm2\pi/3$ in Eq.~(\ref{eq1}) and in fig.~\ref{fig:circle}, which
consequently has a $2\pi/3$ symmetry. This finally leads to a
specific frustration in orbital sector even in simple lattices
such as cubic ones, e.g.\ in
perovskites~\cite{Kugel-Khomskii,Khomskii-Mostovoy}.

The $t_{2g}$ orbitals, e.g. the one shown in
fig.~\ref{fig:orbitals}(c), have two specific features
differentiating them from $e_g$ ones:

(1)~In contrast to $e_g$ orbitals for which the real relativistic
spin-orbit coupling $\lambda\bf l\cdot S$ is in the leading order
absent ($e_g$-states $|z^2\rangle = |l^z{=}0\rangle$,
$|x^2-y^2\rangle = \frac{1}{\sqrt{2}}\{ |l^z {=}{+2}\rangle +
|l^z{=}{-2}\rangle \}$, and the orbital moment is quenched), it is in
general nonzero for $t_{2g}$ states.

(2)~The shape of $t_{2g}$ wave functions is such that e.g.\ in 3d
lattices they can give rise to 2d and even 1d bands. Thus in
perovskite lattices $xy$-orbitals have significant overlap and
hopping only in $xy$-planes, but practically negligible overlap in
$z$-direction, see fig.~4, and as a result the corresponding
tight-binding bands would be two-dimensional, with the dispersion
in $k_x$, $k_y$, but not $k_z$. Even more drastic consequences can
we have in other lattices. Thus the corner-sharing TM tetrahedra
of B-sites in spinels (topologically equivalent to a pyrochlore
lattice), shown in fig.~5, have significant direct overlap of say
$xy$ orbitals along metal chains in $xy$-planes, $xz$ orbitals ---
along $xz$-chains, etc. In effect we would have a collection of 1d
bands. This feature (reduced dimensionality) on one hand can
lead to specific ordering phenomena like Peierls
transition~\cite{Khomskii-Mizokawa}, see below; and, on the other
hand, they can strongly enhance quantum effects in orbital sector
(\cite{Khaliullin-Maekawa}, see also~\cite{van den Brink} and
references therein). Note that in principle one of $e_g$-orbitals,
$x^2-y^2$, also has this property (almost no overlap in the third,
$z$ direction), thus the occupation of this type of orbital can
also lead to a two-dimensional band (this is crucial for High-$T_c$
cuprates), although these effects are more pronounced for $t_{2g}$
systems.

\begin{figure}[htbp]
\centering
\includegraphics[width=7cm]{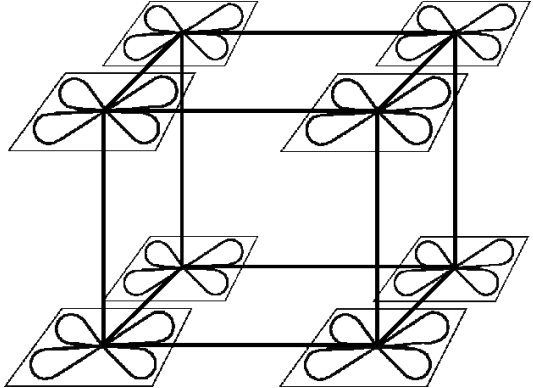}
\caption{\label{fig:xy} The $xy$-orbitals in a simple cubic (or
perovskite) lattice. One can see that there exists strong overlap
in $xy$-plane but practically no overlap in the $z$-direction.}
\end{figure}

\begin{figure}[htbp]
\centering
\includegraphics[width=10cm]{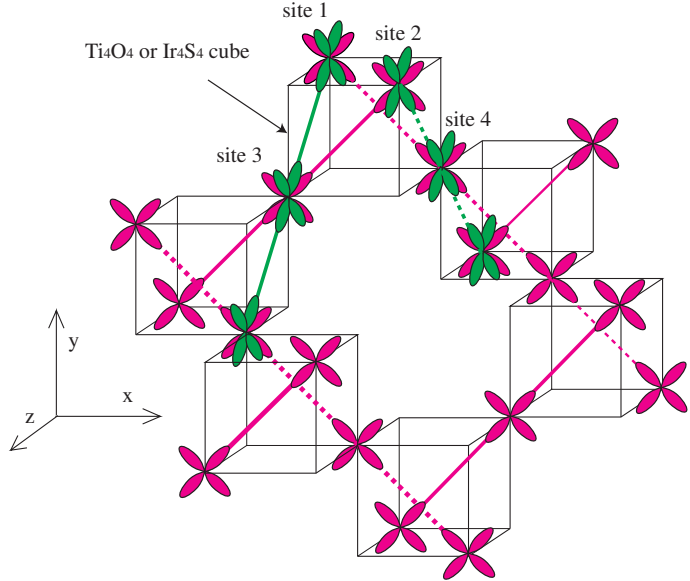}
\caption{\label{fig:spinel}  Overlap of different
$t_{2g}$-orbitals in the $B$-sublattice of spinels (e.g. in
MgTi$_2$O$_4$ or CuIr$_2$S$_4$), showing the formation of
one-dimensional bands.}
\end{figure}

Specific role of orbitals in TM compounds is largely connected
with the famous Jahn-Teller (JT) theorem (which, as Teller himself
wrote in the preface to the book~\cite{Englman}, was actually
suggested to him by Landau).  In a simplest form sufficient for
our purposes it states that the high-symmetry state with an
orbital degeneracy is unstable with respect to a spontaneous
decrease of symmetry lifting this degeneracy (we ignore here
specific quantum ``vibronic'' effects~\cite{Englman,Bersuker}
which can be very important for isolated JT impurities, but which
are less significant for concentrated systems which we consider
here). This spontaneous lifting of symmetry leads to an occupation
of particular orbitals (an orbital ordering (O.O.)), and
simultaneously to a structural phase transition with the reduction
of symmetry (cooperative Jahn-Teller effect). One can not exist
without the other, and it is a definite misunderstanding when
people sometimes are trying to discuss these phenomena as two
separate events.\footnote{There exist different mechanisms of an
O.O. and of the corresponding structural distortion: it may be
just the electron--lattice, or JT interaction (interaction of
orbitals with lattice distortions~\cite{Gehring}), or it can be a
purely electronic (exchange) interaction~\cite{KK,Kugel-Khomskii},
or even a direct quadrupole--quadrupole interaction (relevant for
similar phenomena in rare earths compounds). But in any case, even
if the main driving force of an O.O. is purely electronic, of
course the lattice would react and there will appear a
corresponding lattice (JT) distortion.}

Another question is, which particular effect are we probing by
one or another experimental technique. Different techniques are
more sensitive and may be predominantly determined  either
directly by an orbital occupation, or by the corresponding lattice
distortion. Thus e.g.\ the resonant X-ray scattering at the K-edge
($1s-4p$ transition)~\cite{Murakami} is apparently mostly
determined by lattice distortions~\cite{Elfimov,Joly,Altarelli},
although there may exist in principle also a direct electronic
contribution~\cite{Ishihara-Maekawa}, apparently weaker in this
channel. At the same time the $L_{2,3}$ absorption
($2p-3d$) directly probes an orbital occupation. But in no way
does it mean that these (orbital ordering and Jahn-Teller
distortion) are two different phenomena and that one can exist
without the other, or that they can have e.g.\ different
temperature dependence; this is definitely never the case.

After this general short introduction I will discuss several (not
all, of course) specific situations and phenomena in which
orbitals apparently play an important role. I will start with the
rather well known cases.

\section{Orbital ordering in insulators with \newline ``simple'' lattices}

Typical examples of orbital effects in TM compounds are met in
systems with one electron or one hole in the doubly degenerate
$e_g$ orbitals --- the system containing Mn$^{3+}$ ($t_{2g}^3e_g^1$),
Cr$^{2+}$ ($t_{2g}^3e_g^1$), Cu$^{2+}$ ($t_{2g}^6e_g^3$), low-spin
Ni$^{3+}$ ($t_{2g}^6e_g^1$). These ions give rise to a strong JT
effect, and all of them have a low-symmetry ground state with an
O.O., in which an orbital degeneracy is lifted.  The best known examples
are the colossal magnetoresistance manganites, with the
prototype material LaMnO$_3$, or many Cu$^{2+}$ compounds
including High-$T_c$ cuprates.

Which mechanism is responsible for an O.O. in these systems, is still
not completely clear. Evidently an electron--lattice (JT)
interaction~\cite{Gehring,Englman} is rather strong there. However the
electronic (superexchange)
mechanism~\cite{KK,Kugel-Khomskii} usually also leads to the same
type of orbital ordering as the JT mechanism, and many ab-initio
calculations~\cite{Anisimov-Lichtenstein,Anisimov-Korotin}
reproduce this O.O. even without lattice distortion. The
relaxation of the lattice decreases the energy still further, but
according to these calculations already purely electronic
mechanism gives about 60\% of the total energy gain.

The problem is that typically both the mechanisms, JT and the
electronic one, give rise to the same structure. To evaluate the
relative importance of one or another mechanism of an O.O., it
would be very helpful to find the cases where these mechanisms
would stabilise different states. One such possibility is
discussed in~\cite{insulators}. A more detailed discussion of these
``classical'' cases of O.O. one can find in the general
references~\cite{Kugel-Khomskii,Tokura-Nagaosa,van den Brink}
cited above.

An important question is how one can get the information about an
orbital occupation and O.O. Until recently the main, and
practically the only experimental method to find out an orbital
occupation was the study of crystal structure: by measuring the
local distortion of MeO$_6$ octahedra one could get pretty
reliable information about the detailed type of orbital occupation
at a particular site. Typical situation is the local elongation of
ligand octahedra,\footnote{Although in the simplest treatment for
strong JT ions (partially filled $e_g$-levels) both local
elongation and contraction of ligand octahedra are equivalent, in
practice it is not the case: out of hundreds systems with such JT
ions with localized electrons there are practically none with
locally compressed octahedra. This can be explained by
higher-order coupling and by the anharmonicity effects~\cite{van
den Brink and Khomskii comment to PRL}. Thus one has to be very
careful with the claims sometimes made in theoretical papers,
which consider e.g.\ an alternation of elongated and compressed
octahedra. Note that the situation may be different in systems
with delocalized electrons and with partially-filled bands: these
bands may well be formed by the ``flat'' orbitals like $x^2-y^2$,
with corresponding net tetragonal contraction of the sample. This
rule is also not true for $t_{2g}$ electrons, for which both signs
of distortions (and also trigonal distortions of different sign)
are possible in different situations, see
e.g.~\cite{Kugel-Khomskii,t2g}.} although in general distortions
may be more complicated, e.g. containing    three types of Me$-$O
distances: two long, two intermediate and two short ones. From
these data one can get the type of occupied
orbitals~(\ref{eq1})~\cite{Kanamori,Goodenough}. If  we denote
three Me-O distances as $l$, $m$ and $s$, the  angle $\theta$
characterizing the orbital state (1) is given by the expression
\begin{equation}
\tan(\theta)=\frac{\sqrt{3}(l-s)}{2m-l-s}, \label{eq2}
\end{equation}
or by the corresponding equation with the change $\theta
\rightarrow \theta \pm \frac{2\pi}{3}$.

The other methods traditionally used to study orbital occupation
are those using ESR, and an indirect
information about an O.O. one can obtain from  magnetic properties of
corresponding systems. Also spectroscopic studies are rather
informative in this respect (``ligand field spectroscopy'').

An important  new development is connected with the use of the
resonant X-ray scattering (RXS), initiated by Murakami et
al.~\cite{Murakami}.  As mentioned above, depending on the
specific version of the method used, one again can be more
sensitive to corresponding distortion~\cite{Elfimov,Joly}, but
there are also the ways to probe orbital occupation directly, see
e.g.~\cite{Elfimov,Hatton}. One has already studied by this method
many systems containing Mn, Cu etc. One new and puzzling
phenomenon was observed in many of these studies; in many cases
the intensity of a signal which was attributed to an O.O., changed
rather strongly below magnetic ordering temperature, even in cases
where the O.O. occurs at much higher temperatures, so that it has
to be already saturated at $T_N$. This behaviour was found in
manganites~\cite{Murakami} in which this increase of intensity below $T_N$
is about 30\%, and in KCuF$_3$~\cite{Paolasini KCuF3}, where the
effect is even much stronger: the signal has increased at around
$T_N\sim 30\,\rm K$ by a factor~2.5, despite the fact that an O.O.
exists (and is practically constant) in this system up to the
temperature of its melting or decomposition. What is the
explanation of this effect, is completely unclear at present. One
possibility is that there may be a direct contribution of magnetic
ordering (via spin-orbit coupling or via interference of different
scattering channels) to the signal which was
attributed to an O.O\null.  Whether this possibility may be
realised in practice, is an open question.

To finish with the discussion of the conventional effects
connected with an O.O., one should mention that it largely
determines the character of an exchange interaction and the type
of magnetic ordering in corresponding systems. This is the essence
of the famous Goodenough--Kanamory--Anderson (GKA) rules, see e.g.\
\cite{Goodenough,Khomskii in Spin Electronics}. In short, the main
ones of them are: there is a strong antiferromagnetic coupling if
on corresponding sites the (singly-) occupied orbitals are
directed towards each other. If however an occupied orbital is
directed towards an empty (or doubly-occupied) one, there will be
a weaker ferromagnetic coupling.

The interplay between an orbital occupation  and magnetic ordering
in principle gives rise to the possibility of a change of an O.O.
at the magnetic transition, but for $T_N \ll T_{\rm O.O.}$  this
effect is apparently too weak to explain the anomalies at $T_N$ in
RXS discussed above.

\section{Reduced dimensionality due to orbital ordering}
Specific feature of orbital degrees of freedom is an anisotropy of
corresponding electron distribution. Consequently, particular
orbital occupation can lead to the appearance of a strong anisotropy
in the properties of such systems, even if the original
crystal structure is relatively isotropic.

There are many examples of this phenomenon. Thus, in undoped
manganites with perovskite structure the magnetic ordering is of
layered type (A-type ordering --- ferromagnetic planes stacked
antiferromagnetically)~\cite{Wollan}. Even more striking is the
example of KCuF$_3$: in this practically cubic crystal, due to an
orbital ordering (alternation of hole orbitals $x^2-z^2$ and
$y^2-z^2$) magnetic properties are those of a
quasi-one-dimensional
antiferromagnet\cite{KCuF3,KK,Kugel-Khomskii}. Despite cubic
lattice, this material is one of the best one-dimensional
antiferromagnets.

Orbital ordering can also strongly modify electronic structure and
properties of some systems, such as conductivity. Thus,
in certain doped manganites, e.g.\ in Nd$_{1-x}$Sr$_x$MnO$_3$ for
$x\sim0.6$, predominantly $x^2-y^2$ orbitals are occupied, forming
corresponding partially filled band. Consequently this material
has much higher conductivity in the $xy$-plane than perpendicular
to it. Apparently two-dimensionality of most High-$T_c$ cuprates,
although largely due to their layered structure, is substantially
enhanced by the location of charge carriers (here holes) mostly in
$x^2-y^2$ bands.

Yet another recent example is the new spin-Peierls system
TiOCl~\cite{TiOCl}: due to a particular orbital occupation this
material, that structurally is a layered system, electronically
becomes quasi-one-dimensional, which finally leads to a
spin-Peierls transition in it.

\section{Orbitally-driven Peierls state}
\begin{figure}[htbp]
\centering
\includegraphics[width=10cm]{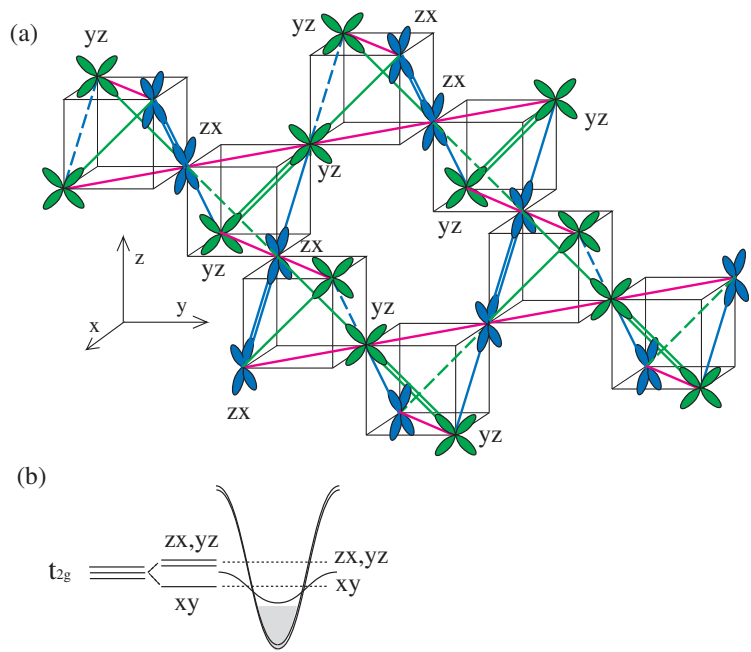}
\caption{\label{fig:MgTi} Schematic band structure and orbital ordering in
MgTi$_2$O$_4$ leading
to the formation of ``chiral" superstructure (by
\cite{Khomskii-Mizokawa})}
\end{figure}

\begin{figure}[htbp]
\centering
\includegraphics[width=10cm]{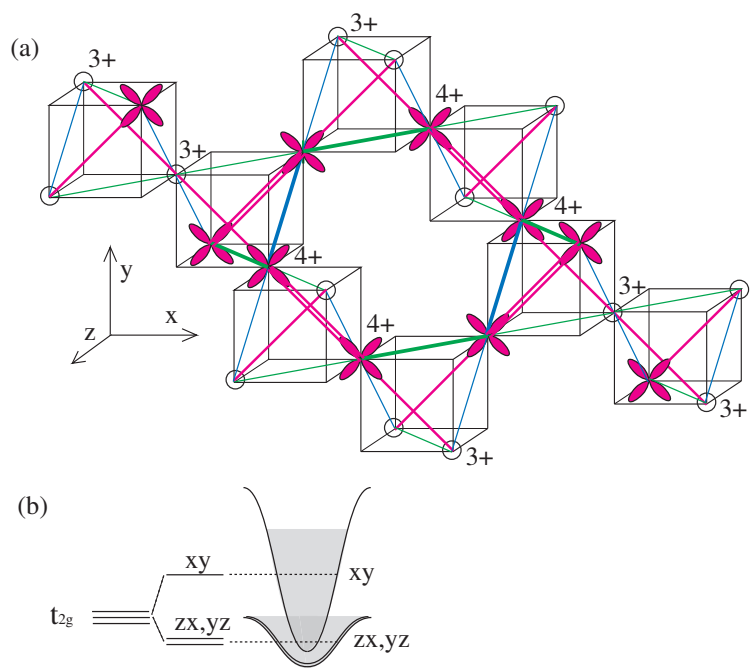}
\caption{\label{fig:CuIr} Schematic band structure and orbital ordering in
CuIr$_2$S$_4$ leading
to the formation of octamers (by \cite{Khomskii-Mizokawa})}
\end{figure}

TiOCl is not the only example of quasi-1d behaviour caused by
orbital ordering. Even more striking example is provided by some
spinels with TM on B-sites and with partial occupation of
$t_{2g}$-levels. In some of such systems quite spectacular
superstructures were observed recently: an ``octamer'' ordering in
CuIr$_2$S$_4$~\cite{Radaelli} or ``chiral'' structural distortion
in MgTi$_2$O$_4$~\cite{Schmidt}. The natural explanation of such
strange structures can be found in the concept of an
orbitally-driven Peierls state~\cite{Khomskii-Mizokawa}. As
illustrated in fig.~5, in this crystal structure different
$t_{2g}$ orbitals overlap only with corresponding orbitals of
neighbouring sites along particular directions: $xy$-orbital with
$xy$ along $xy$-chain, $xz$ with $xz$ in $xz$-chains, etc. As a
result the electronic structure consists in a simplest
approximation of three degenerate 1d-bands. In both CuIr$_2$S$_4$
and MgTi$_2$O$_4$ there occurs with decreasing temperature a
metal--insulator transition with the cubic--tetragonal lattice
distortion. This distortion splits three degenerate bands, so that
either only one or two of them are partially occupied, figs.6(b),
7(b). In both cases the corresponding bands turn out to be
$\frac14$ or $\frac34$-filled, and as a result the system
undergoes a Peierls-like transition --- {\it tetramerization}
along respective directions. It is driven by orbital ordering (one
may call it an ODW --- Orbital Density Wave). Simultaneously it
also leads to a formation of spin singlets on dimers with orbitals
directed towards one another (double bonds in figs.6(a), 7(a)).
One can see that the resulting superstructures exactly coincide
with those observed experimentally in~\cite{Radaelli,Schmidt}:
they give chiral superstructure in MgTi$_2$O$_4$~  and octamers in
CuIr$_2$S$_4$ \cite{Khomskii-Mizokawa}, see figs.~\ref{fig:MgTi}
and \ref{fig:CuIr}.

One can argue that similar phenomenon should occur also in some
other systems, e.g.\ in NaTiO$_2$~\cite{Khomskii-Mizokawa}. It may
be also relevant for the transition to a spin-gap state in
La$_4$Ru$_2$O$_{10}$~\cite{Khalifah}, and possibly even to an old
problem of Verwey transition in
magnetite~\cite{Verwey,Khomskii-Mizokawa}. Thus, in
La$_4$Ru$_2$O$_{10}$ the appearance of a singlet ground state
below structural phase transition was originally interpreted
in~\cite{Khalifah} as a transition of each Ru$^{4+}$ ($d^4$) from
$S=1$ ion into a nonmagnetic $S=0$ state. However this seems to be
rather unlikely, as it would require the splitting of $t_{2g}$
levels larger than the Hund's rule coupling $J_H$ which for $Ru$ is of
order $0.6\,\rm eV$. Most probably singlet states in this system
are again those on Ru dimers, stabilized by corresponding orbital
ordering. This picture is supported by recent LDA${}+{}$U
calculations~\cite{HuaWu}, and it seems to agree with the results
of recent inelastic neutron scattering~\cite{R.Osborn}.

An example of a more complicated singlet Peierls-like state is
provided by LiVO$_2$ --- the system with a quasi-two-dimensional
triangular lattice, fig.~8. A structural transition accompanied by
an opening of a spin gap in this system may be explained by an
orbital ordering with the formation of three orbital sublattices
in it~\cite{H.Pen}, of the type $(xy,xz)$, $(xy,yz)$ and
$(xz,yz)$, see fig.~9. One sees that as a result of this ordering
there will appear strong antiferromagnetic coupling in some
triangles of V (shaded triangles in fig.~9); this would lead to
the formation of singlets on these triangles (three spin 1
V$^{3+}$ ions combine into a singlet). One can say that this is a
``next level of complexity'' --- singlets not on dimers, as in Peierls
or spin-Peierls case, but on trimers.

\begin{figure}[htbp]
\centering
\includegraphics[width=7cm]{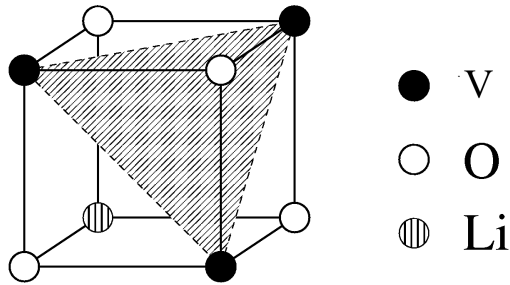}
\caption{\label{fig:LiVO2} Crystal structure of
LiVO$_2$, showing the formation of quasi-two-dimensional triangular lattice of
magnetic ions $V^{3+}$ ($t_{2g}^2$).}
\end{figure}

\begin{figure}[htbp]
\centering
\includegraphics[width=7cm]{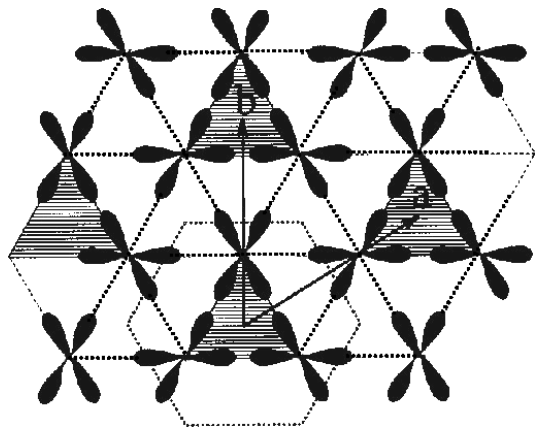}
\caption{\label{fig:triangles} Three-sublattice ordering of two occupied
$t_{2g}$-orbitals of $V$ in LiVO$_2$ \cite{H.Pen}. Shaded are the spin
singlet triangles.}
\end{figure}

\section{Possible role of orbitals in insulator-metal transitions}
In previous sections we saw that orbital degrees of freedom play
an important role in the IMT and in corresponding structural
modifications in some spinels. However one can argue that this factor has
broader significance, and presumably orbitals play an important
role in IMT in many other systems as well. The study of this
question is now only at the beginning, thus my discussion here
would have mostly a qualitative character and would rely only on a
few examples.

Among  the best known systems with IMT are vanadium oxides, notably
V$_2$O$_3$ and VO$_2$, see e.g.~\cite{Fujimori RMP}. As to
V$_2$O$_3$, the idea of the importance of an O.O. in it was suggested
long ago~\cite{Castellani} and was revived in~\cite{Rice} to
explain neutron scattering results~\cite{Aeppli-Broholm} that the
magnetic correlations in the metallic phase are quite different
from those expected from the long-range magnetic order in the
insulating phase. O.O was also invoked initially to explain the
results of RXS in V$_2$O$_3$~\cite{Paolasini}. And although the initial model
of~\cite{Castellani} is apparently faulted~\cite{Ezhov V2O3}, and
the results of~\cite{Paolasini} can be explained without invoking
O.O.~\cite{Lovesey}, still direct X-ray absorption
measurements~\cite{Park-Tjeng} show that indeed there is a change
of an orbital occupation accompanying IMT, although it is much
weaker than the one considered in~\cite{Castellani,Rice,Paolasini}.

The case of VO$_2$ is more interesting from our point of view. The
IMT in VO$_2$, occurring at about $70\,\rm C$, is accompanied by
the structural transition from the rutile (R) to the monoclinic (M1)
phase, in which V chains along $c$-axis of  R structure are
dimerized, and simultaneously there occurs a ``twisting'' ---
tilting of V dimers, caused by the antiferroelectric-type shifts
of V ions towards oxygens in $ab$-plane. As suggested already long
ago by Goodenough~\cite{Goodenough VO2}, this ``twisting'' shifts
and depopulates antibonding $\pi^*$-band made of V orbitals lying
mostly in $ab$-plane, so that only the 1d band made of one type of
$t_{2g}$ orbitals of V, which have strong direct overlap along
$c$-chains, is occupied. This 1d half-filled band, in its turn,
leads to a  Peierls distortion (dimerization of $c$-chain), and
finally the material becomes insulating.

We see that in this picture the change of orbital occupation at
the IMT plays crucial role in the transition itself. This picture
is now confirmed by a number of ab-initio calculations~\cite{VO2
bands}, and very recently it was proven by the direct X-ray absorption
measurements~\cite{Haverkort VO$_2$}. Thus at least in VO$_2$ the
orbital reorientation is extremely important for the IMT.

One can even qualitatively understand why orbital effects are
stronger in VO$_2$ than in V$_2$O$_3$. There exist a strong
tendency of $d^1$ ions to form spin singlets in many TM compounds.
Besides VO$_2$, similar phenomenon is also observed in
Ti$_2$O$_3$~\cite{Goodenough} and in the Ti Magneli phases,
e.g.~in Ti$_4$O$_7$~\cite{Schlenker Ti4O7}. But this tendency is
much less prominent for $d^2$, $d^3$, etc, configurations. And
apparently the tendency to form singlet pairs is greatly enhanced
by corresponding orbital ordering: it is most favourable for that
to put one electron at each site of the pair into orbitals
directed towards one another.

As I said, the  question of the role of orbitals at IMT's is just
started to be investigated, and it is not a priori clear how
important is this factor in general; it may indeed be
system-dependent. But one can give some arguments that it can help
to explain one general open problem which seems to be quite common
for many IMT's in TM compounds and  which did not attract yet
sufficient attention:

In most of the TM systems with IMT  the energy gap $E_g$ which
opens in the insulating phase is much larger than the
corresponding $T_c$. Thus e.g.\ in V$_2$O$_3$ and in VO$_2$ $E_g$
is $\sim0.5$--$0.6\,\rm eV$, whereas $T_c$ is respectively
$150\rm\,K$ and $370\,\rm K$; in Fe$_3$O$_4$ $E_g\sim0.3\,\rm eV$
and $T_c=119\,\rm K$, etc.  This large discrepancy is not an
exception but rather a rule for  IMT in this class of compounds.
This is in  strong contrast
to the situation  e.g.\ in superconductors or in systems with CDW
or SDW, where one typically has the values of the gap of the order
of the theoretical ones for BCS superconductors, $2\Delta/T_c =
3.5$. Such values we should expect also for Peierls transitions
(which is actually the case in several low-dimensional
materials~\cite{1d}) or in spin-Peierls
systems~\cite{spin-Peierls}. Apparently this large discrepancy at
IMT in TM compounds is a consequence of strong electron
correlations in these systems, and it can serve as a signature of
their importance.

However {\it why} is it the case, is not really clear. In simple
treatments of the Mott-Hubbard transitions, to get small $T_c$
one  would require an extremely fine tuning of the parameters
(electron hopping $t$ or bandwidth $W=2zt$ and the Hubbard
repulsion~$U$) which determine whether the material would be a
metal or an insulator: typically both $W$ and $U$ are of order of
several eV, and one indeed needs their almost exact cancellation
to get $T_c$ of order of $100\,\rm K$, or $0.01\,\rm eV$. More
sophisticated calculations, e.g.\ DMFT~\cite{Kotlyar  RMP},
sometimes (re)produce small $T_c$ and large values of $E_g/T_c$,
although one usually has to invoke some extra factors such as
frustrations. Physical picture used in this approach is that of
``preformed'' gap of order of $U$ (between lower and upper Hubbard
bands), whereas in the metallic phase close to IMT there exists
also a small coherent peak at the Fermi-level. But again, the
widths of this peak which determines the energy scale of IMT is
small only when we have fine tuning of $W$ and~$U$.

One possible factor which can help to resolve this problem is an
eventual change of spin and orbital correlations at the IMT. In
real systems not only does the gap close at the IMT, but also
magnetic and orbital order, or correlations, change significantly.
Above we  discussed the orbital change in VO$_2$. In V$_2$O$_3$
probably the role of orbitals is less important, but, on the other
hand, spin correlations change a lot~\cite{Aeppli-Broholm},
possibly even becoming ferromagnetic above
$T_c$~\cite{Ziebeck-Neumann}.  Orbital modification, e.g.\ in
VO$_2$, can make the effective bandwidth smaller in the insulating
case (simply speaking, for example making the energy bands
one-dimensional instead of three-dimensional ones above $T_c$, or
removing the crossing of different bands). Then one can have the
situation that $W > U$ above $T_c$, but becomes (much) smaller
below it. If so, we would not need such a ``fine tuning'' of $W$
and $U$, as is required in the simple nondegenerate Hubbard model.

Change of spin correlations (in its turn also connected with the
change of orbital occupation)  can also facilitate  strong IMT
with opening of a large gap, because it can lead to a change of an
effective value of the Coulomb (Hubbard) repulsion
$U$~\cite{Park-Tjeng}. Indeed, typically spin ordering or
correlations in the insulating state are  antiferromagnetic. Then
the virtual transition of an electron to a neighbouring site,
determining the value of $U_{\rm eff}$, is just the usual $U$ in
the Hubbard model. However, as is often the case, magnetic
correlations in the metallic phase are smaller or even become
ferromagnetic~\cite{Ziebeck-Neumann}; then such virtual transition
would ``cost'' the energy $U - J_H$ (or several $J_H$ in case of
many-electron ions), where $J_H$ is the intra-atomic Hund's rule
exchange (of order $0.8$--$0.9\,\rm eV$ for 3d ions). Thus
effectively the value of $U$ may strongly decrease above $T_c$,
which can be traced back to the presence and  modification of
orbital occupation at the IMT. This effect again makes the
conditions for the IMT less stringent, and can at least partially
explain large values of energy gaps in the insulating phase.

\section{Orbitals in frustrated lattices}
As I already said, an orbital exchange may be frustrated even in
simple lattices such as square or cubic
ones~\cite{Kugel-Khomskii,Khomskii-Mostovoy}. These frustrations
would become even more prominent in the lattices with geometric
frustrations~\cite{Ramirez}, such as triangular, kagome or
pyrochlore lattices. Frustrated systems attract now considerable
attention, but mostly from the point of view of their spin
properties. However orbitally-degenerate frustrated systems also
present significant interest. The system which was most widely
discussed in this context is LiNiO$_2$, containing low-spin
Ni$^{2+}$ ions with the configuration $t_{2g}^6e_g^1$ on a
two-dimensional triangular lattice, similar to that in LiVO$_2$
shown in fig.~8. This state has both $S=\frac12$ and the double
orbital degeneracy on this lattice. This system was first studied
as one of the best candidates for the Anderson's RVB state, but
later it was suggested that there exists in it also an orbital
liquid state~\cite{Kitaoka}. Experimentally indeed there exist in this
system  neither  long-range magnetic order, nor structural
transition required by orbital degeneracy.

Theoretical treatment of this situation~\cite{LiNiO2 we} has shown
that there exists a large orbital degeneracy in this system,
which however can be lifted by the order-from-disorder
mechanism~\cite{Villain,Shender}. Magnetic interactions, however,
were shown to be predominantly ferromagnetic in the Ni layer.
Similar system NaNiO$_2$ indeed shows the behaviour obtained
theoretically: there exists in it a structural transition with an
O.O. at $T_{\rm str} = 480 K$, and at lower temperatures - a magnetic
ordering with
ferromagnetic layers coupled antiferromagnetically~\cite{Grenoble
NaNiO2}. The absence of a long-range magnetic ordering in
LiNiO$_2$ may be explained by interlayer
frustrations~\cite{Grenoble,LiNiO2 we}, which is probably responsible for the spin-glass transition observed in LiNiO$_2$ at about $8\rm\,K$~\cite{Grenoble}). However  the absence of
JT transition in it is not yet explained. (Recently a short-range
orbital order in LiNiO$_2$ was detected by EXAFS~\cite{EXAFS in
LiNiO2} and by PDF~\cite{Egami}).

\section{Some extra remarks}
In this short article I described certain phenomena in which
orbital degrees of freedom play an important role. At the end I
want also to make a few other, somewhat speculative comments about
certain other possible manifestations of orbitals in the structure
and properties of systems with strongly correlated electrons.

1) One may argue that the very formation of certain crystal
structures is at least partially determined by orbital degrees of
freedom. The best known example is probably given by many
Cu$^{2+}$ compounds. It is well known that because of the
extremely strong JT effect,  Cu$^{2+}$ always exists either in a
strongly distorted (elongated) ligand octahedron, or this
elongation is so strong that one or two apex ligands ``go to
infinity'', leaving Cu$^{2+}$ in a 5-fold pyramid or 4-fold square
coordination. These coordinations are very typical for Cu$^{2+}$
and for many compounds containing it, including such important
ones as High-$T_c$ cuprates. The very existence of e.g.\ YBCO
superconductor structure  is largely connected with this factor.

2) Another, less clear but rather suggestive case, is the structure
of hexagonal manganites $R$MnO$_3$, $R={}$small rare earth (RE) or
Y and Sc.\footnote{These systems attract now considerable
attention because they are one of the best known example of
multiferroics --- materials that combine magnetic ordering with
ferroelectricity~\cite{multiferroics}} Mn$^{3+}$ ions in these systems
are 5-fold coordinated (located in the centre of oxygen trigonal
bipyramid). Interestingly, Mn is the only TM element forming this
crystal structure: all the others, including e.g. Fe$^{3+}$
($d^5$) (orthoferrites)  or Cr$^{3+}$ ($d^3$) (orthochromites)
form (distorted) perovskite structure even for small~RE. Why is
that, is not completely clear, but one factor may be that
Mn$^{3+}$ in an octahedral coordination, typical for perovskites,
is a strong JT ion. Probably the combination of JT distortion with
strong tilting, required for small RE, is not very favourable
(although one can still stabilize RMnO$_3$ with small RE in a
perovskite structure). Thus it is feasible that, instead of trying
to lift the JT degeneracy, the system simply chooses another crystal
structure --- that of hexagonal YMnO$_3$, in which this degeneracy
is absent. Indeed, the CF splitting of 3d-levels in trigonal
bipyramid coordination is into two doublets and an upper singlet,
and 4 electrons of Mn$^{3+}$ occupy two lowest doublets, so that
no orbital degeneracy is left.

3) Yet another, also rather speculative example of a possible role of
orbital degeneracy in apparently unrelated phenomena may be
met in TM compounds in which there exist a spontaneous charge, or
valence disproportionation. An example of this phenomenon is given
e.g.\ by ferrates like CaFeO$_3$~\cite{CaFeO3}, in which there
occurs charge disproportionation
2Fe$^{4+}{}\to{}$Fe$^{3+}{}+{}$Fe$^{5+}$ (of course, this process
is never complete, and actually the charge modulation is much less
than the one that would follow from this formula, but the quantum numbers of
the resulting states indeed coincide with those of Fe$^{3+}$ and
Fe$^{5+}$).

Another system in which similar phenomenon apparently takes place
is perovskite nickelates RNiO$_3$ with the low-spin
Ni$^{3+}$~\cite{Mizokawa-Khomskii}. The appearance of two
inequivalent Ni's was established at least for small RE and
Y~\cite{Garcia-Munoz}, and possibly in all these systems there
occurs charge disproportionation of the type
2Ni$^{3+}{}\to{}$Ni$^{2+}{}+{}$Ni$^{4+}$ (another option is that
both Ni ions are Ni$^{2+}$, but the extra hole is located at every
second {[111]} layer of oxygens~\cite{Mizokawa-Khomskii}).

Interestingly enough, in both these cases the starting,
homogeneous state would correspond to the situation with orbital
degeneracy: high-spin Fe$^{4+}$ ($t_{2g}^3e_g^1$) and low-spin
Ni$^{3+}$ ($t_{2g}^6e_g^1$) are both strong JT ions. It is not
actually clear if this factor is really important in causing
charge disproportionation, but one may argue  that this
disproportionation is one way to get rid of orbital degeneracy:
instead of doing it via JT distortion, the system does it by simply
getting rid of the degenerate electron! (the resulting states
Fe$^{3+}$, Fe$^{5+}$, or Ni$^{2+}$, Ni$^{4+}$ are all nondegenerate).

There are many other effects connected with orbital degrees of
freedom in TM compounds. I can not dwell on all of them here and
will only list some of them with short comments and some
references:

({\it a})~Apparently orbitals play an important role in double
exchange in manganites and similar materials. Thus, their
inclusion helps to explain the absence of ferromagnetism and the
appearance of unusual magnetic structures in overdoped manganites
--- leading to a marked asymmetry in their properties for
underdoped (hole-doped) and overdoped (electron-doped)
systems~\cite{van den Brink and Khomskii}

({\it b}) In connection with the colossal magnetoresistance manganites,
the question arises
what is the orbital state and the role of orbitals in ``optimally
doped'' ferromagnetic metallic state. Experimentally at low
temperatures there are no indication of JT distortion, even
local~\cite{Egami-Louca}. One possible explanation is that here we
are dealing with an orbital liquid, stabilized by
doping~\cite{Ishihara-Nagaosa}. However also a more exotic
possibility was discussed in this context~\cite{complex} --- that with ordering
of {\it complex} orbitals, of the type of~(\ref{eq1}) but with
complex coefficients, e.g.
\begin{equation}
(|z^2\rangle + i|x^2-y^2\rangle)/\sqrt2\,.
\end{equation}
This state has cubic symmetry and causes no lattice distortion,
but it has a magnetic octupole moment.

({\it c})  Orbitals may play a role in charge ordering (CO), often
observed in doped TM compounds. One such example, that of
CuIr$_2$S$_4$, was already mentioned above. Possibly O.O. is also
relevant for the low-temperature behaviour of magnetite: recent
LDA${}+{}$U calculations~\cite{Leonov} gave such O.O. for the
crystal structure obtained in~\cite{Radaelli-Attfield}.

({\it d})  In connection with the problem of CO, one has to
mention the possibility of two types of it: the conventional
site-centered ordering, and a bond-centered one (called Zener
polaron state in~\cite{Daoud}). Charge/octamer ordering in
CuIr$_2$S$_4$ can be viewed both as a site-centered  charge and
orbital ordering and as a bond-centered formation of spin singlets
at certain bonds; the same is true for the insulating state of
VO$_2$ and in Magneli phases of Ti and V\null. Similar coexistence
of  site-centered and bond-centered CO, facilitated by
corresponding O.O., apparently may be present in slightly
underdoped manganites, and the resulting state may be
ferroelectric --- a new mechanism of ferroelectricity in magnetic
systems~\cite{Efremov}. This factor can also be important in the
low-temperature phase of magnetite Fe$_3$O$_4$, and it can explain its
multiferroic behaviour~\cite{DKh}.

\medskip
In this mini-review I tried to show that the orbital degrees of
freedom, especially in case of orbital degeneracy, give rise to
multitude of consequences. Many of them are already well known,
but this field is definitely far from closed and still produces
new and new  surprises.

\medskip
I am very grateful to many colleagues for numerous discussions of
these problems, but especially to J.~van~den~Brink, K.~Kugel,
T.~Mizokawa,\break M.~Mostovoy, G.~A.~Sawatzky and L.~H.~Tjeng. This
work was supported by the Deutsche Forschungsgemeinschaft via SFB 608.

\def\bbf#1,{{\bf #1},}

\end{document}